\documentstyle[epsfig]{aipproc}

\begin{document}

\rightline{KFA-IKP(TH)-97-12}

\title{The BFKL-Regge Expansion\\
 for the Proton Structure Function \\
at Small $x$}
\author{Vladimir  R. Zoller}
\address{Institute for Theoretical and Experimental Physics\\
 117218 Moscow, Russia {\thanks  
{Talk at 5th International Workshop on Deep Inelastic Scattering
and QCD (DIS'97), Chicago, April 1997}}}

\maketitle

\begin{abstract}
We report an evaluation of 
  subleading eigenvalues 
 and eigenfunctions
of the BFKL equation in the color dipole representation
with a running gauge coupling.
We present an expansion of the  small-$x$ proton structure function $F_{2p}(x,Q^2)$
  in terms of
the  rightmost BFKL-singularities.
The  BFKL-Regge 
 phenomenology of  DIS structure functions 
 is developed
which is shown to provide remarkably good
description of the data on  
 $F_{2p}(x,Q^2)$ from E665 to HERA.

\end{abstract}
\vspace{-0.8cm} 

\subsection*{Introduction}
  In this talk I address  the issue 
 of extrapolation of the proton structure functions
to a domain of very small $x$ from the attainable 
kinematical region of 
 $x$ and $Q^2$. So, this is an old problem 
  of predicting  the future from the known past.
The standard predictor
 is the  GLDAP 
 evolution
equation \cite{GRIB72}.
However, the art of predicting is difficult.  
The  numerous 
pre-HERA  GLDAP fits to the proton structure functions 
 are
equally good for large-$x$-data but seldom come close to the
new data points at smaller $x$. The lack of 
predictive power is not surprising and we comment on this point below.
 It is well known that the 
GLDAP evolution breaks down at small $x$ and  is superseded
by the $\log(1/x)$-BFKL-evolution \cite{LIPAT76}. Contrary to the 
GLDAP approach, the BFKL evolution  predicts uniquely 
 the proton structure functions at 
an arbitrarily small $x$  from   the  input
 at a starting point  $x=x_0$.
 DIS structure functions at 
asymptotically large $1/x$
are dominated by the rightmost Regge-singularity
with the intercept $\Delta_0$ \cite{LIPAT86}  
 and
\begin{equation} 
F_{2p}(x,Q^2) = F_2^{(0)}(Q^2)\left(1\over x\right)^{\Delta_0}\,. \label{eq:F20}
\end{equation}
At moderate $x$, however,  the  subleading contributions 
 to $F_{2p}(x,Q^2)$ 
with smaller intercepts $\Delta_1, \Delta_2,...$
can not be neglected.  The task of my talk is to present
an evaluation of the intercepts $\Delta_n$, the pomeron
 trajectory slopes
$\alpha_n^{\prime}$
and the corresponding structure functions $F_2^{n}(Q^2)$,
to arrive finally at the representation 
\begin{equation}
F_{2p}(x,Q^2) =\sum_n  
A_nF_2^{(n)}(Q^2)\left({x_0\over x}\right)^{\Delta_n}\,. \label{eq:REGGE}
\end{equation}
I conclude with the expansion of the proton
structure function $F_2(x,Q^2)$ in terms of the three rightmost
BFKL singularities.
Such a three pole approximation 
 seems to exhaust the existing experimental data  
thus providing a reliable basis  
for the  BFKL-Regge phenomenology of  diffractive 
DIS. 
\subsection*{The BFKL  Eigenvalue Problem in the  Dipole Picture} 
  The virtual photo-absorption cross section $\sigma(r,x)$ , where
$r$ is the  color dipole size, satisfies
 the BFKL equation \cite{PISMA1}
(hereafter $\xi=\log(1/x)$):
\begin{equation}
{\partial \sigma(\xi,r)\over \partial \xi} ={\cal K}\otimes
\sigma(r,x) \label{eq:BFKL}
\end{equation}
with the kernel $\cal K$ which involves the  running
gauge coupling and the infrared cutoff-  
the correlation radius $R_c$ of perturbative gluons.   
We  look for the solution with the Regge behavior 
\begin{equation}
\sigma_n(r,x) =
\sigma_n(r)\left(1\over x\right)^{\Delta_n}\,. \label{eq:SIGMA0}
\end{equation}
Then the eigenfunctions $\sigma_n(r)$ and the eigenvalues $\Delta_n$
are determined from 
\begin{equation}
{\cal K}\otimes \sigma_n(r) =
\Delta_n\sigma_n(r)\,. \label{eq:EIGEN}
\end{equation}
The short-distance asymptotics of the eigenfunctions is known
in the analytic form \cite{PLNZ1}
\begin{equation}
\sigma_{n}(r)=
r^{2}\left[{1\over \alpha_{S}(r)}\right]^{\gamma_{n}-1}\,,
\label{eq:ANAL}
\end{equation}
 where
$\gamma_{n}\Delta_{n}={4/3}$. 
At large distances, $r\gg R_c$,
\begin{equation} 
\sigma_n(r) \equiv \bar\sigma_n= const \label{eq:CONST}
\end{equation}
 due 
to the finite $R_c$. Another useful clue is that
the leading eigenfunction $\sigma_0(r)$ is node 
free and the $ n$-th subleading
solution must have $n$ nodes. Then a practical approach to the 
eigenvalue problem is a variational procedure \cite{NOVIK} applied to a class of
 $n$-node polynomials ${\cal P}_n(z)$ in a variable
$z\sim [1/ \alpha_S(r)]^{\gamma}$.

\subsection*{Results and Discussion}

The BFKL equation with running coupling and infrared cutoff
has a discrete spectrum. 
The  eigenvalues 
$\Delta_n$ obtained by  the variational
method for $n=0,1,2,3,...$ are as follows
\begin{equation}
\Delta_n = 0.40,\,\,0.220,\,\, 0.148,\,\, 0.111,
\,\,0.088,\,\, 0.073,\,\, 0.063,...\,.
\label{eq:DELTAN}
\end{equation}
To an   accuracy
better than $10\%$ the above  series   follows the  
law 
\begin{equation}
\Delta_n={\Delta_0\over (n+1)} \label{eq:DELTA}
\end{equation} 
derived by Lipatov \cite{LIPAT86} from quasi-classical
considerations.

\begin{figure}[b!]
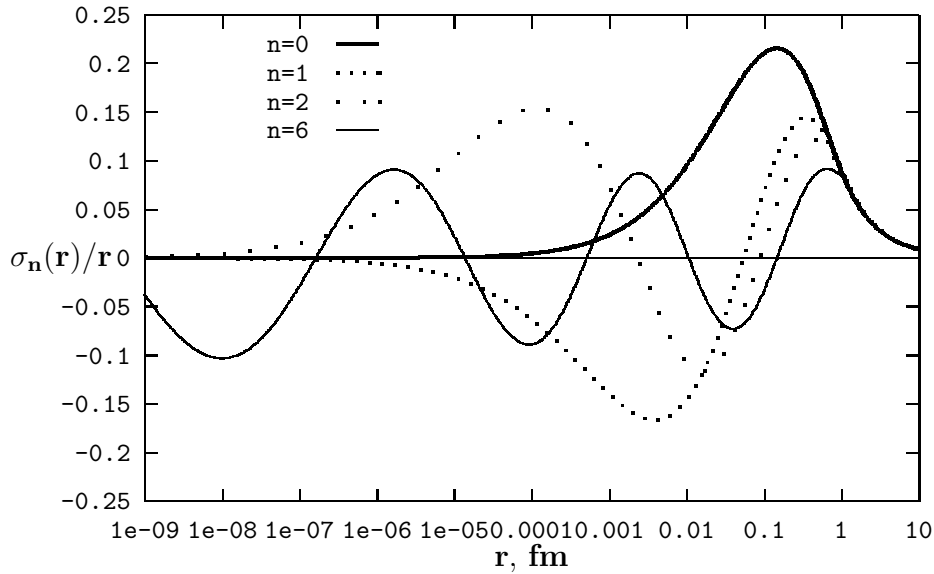
 

\setlength{\unitlength}{0.240900pt}
\ifx\plotpoint\undefined\newsavebox{\plotpoint}\fi
\sbox{\plotpoint}{\rule[-0.200pt]{0.400pt}{0.400pt}}%


\vspace{10pt}
\caption{ BFKL eigenfunctions.}
\label{figvrz1}
\end{figure}

The BFKL eigenfunctions are represented  (Figure \ref{figvrz1})
in term of the  quantity  $\sigma_n(r)/r$ which to a crude 
approximation is similar to Lipatov's quasi-classical eigenfunctions,
which are ${\cal E}_n(r)\sim \cos[\phi(r)]$  for $n\gg 1$ \cite{LIPAT86}.
    
 Once $\sigma_n(r)$ is known, the pomeron trajectory
slope $\alpha_n^{\prime}$ can  readily be derived. In the notations
of ref.\cite{PISMA2}
\begin{equation}
\alpha_n^{\prime}={3\over 32\pi R_c^2}\int d\rho^2\,\rho^2\alpha_S(\rho)
K_1^2(\rho/R_c)\left[1+{\sigma_n(\rho)/ \bar\sigma_n}\right]\,,
\label{eq:ALPHA}
\end{equation}
where (see eq.\ref{eq:CONST})
\begin{equation}
\bar\sigma_n={3\over 2\pi R^2_c}\int d\rho^2 \alpha_S(\rho)
K_1^2(\rho/R_c)\sigma_n(\rho)\,. \label{eq:BARSIG}
\end{equation}
Numerically, $\alpha_0^{\prime}=0.072\, GeV^{-2},\,\alpha_1^{\prime}=0.066\,GeV^{-2},
\,\alpha_2^{\prime}=0.062\, GeV^2,\,\alpha_3^{\prime}=0.060\, GeV^2 $.

\begin{figure}[b!]
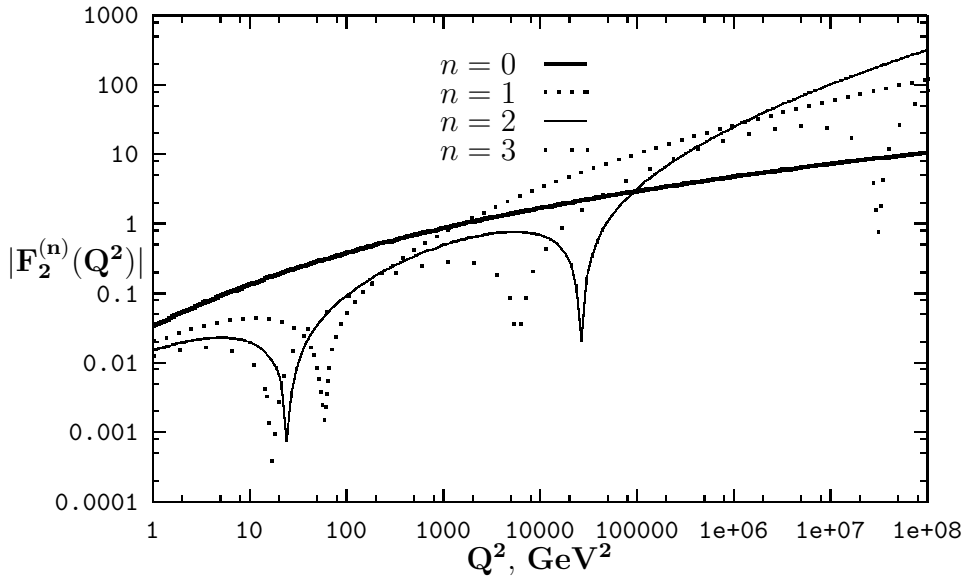
 


\setlength{\unitlength}{0.240900pt}
\ifx\plotpoint\undefined\newsavebox{\plotpoint}\fi
\sbox{\plotpoint}{\rule[-0.200pt]{0.400pt}{0.400pt}}%


\vspace{10pt}
\caption{The modulus of  structure function $F^{(n)}_2(Q^2)$ for $n=0,1,2,3$}
\label{figvrz2}
\end{figure}

The color dipole factorization relates the dipole cross sections
with the structure functions
$F_2^{(n)}(Q^2)$
which are shown in Figure \ref{figvrz2}.
At large $Q^2$, far beyond the nodal region,
\begin{equation}
F_2^{n}(x,Q^2)\propto 
\left(x_0\over x\right)^{\Delta_n}\left[{1\over \alpha_S(Q^2)}\right]^{4/3\Delta_n} \,.
\label{eq:F2ASS}
\end{equation}
Since the relevant variable is a power of the inverse gauge coupling
the nodes are spaced by 2-3 orders of magnitude in $Q^2$-scale
 and only
the first  two of them  are  in the accessible range of $Q^2$.
The first nodes of $F_2^{(n)}(Q^2)$
are located at $Q^2\sim20-60\, GeV^2$. Hence, only the leading
structure function contributes significantly in this region.
 This explains the precocious  BFKL asymptotics found 
 from the numerical solution of the BFKL equation \cite{PLNZ2}.

In the accessible range of $x$ the subleading contributions
are numerically large. In particular, 
the BFKL expansion of the Born approximation dipole cross section
which is used as the boundary 
condition at $x=x_0=0.03$ suggests more than $60\%$ contribution
from $F_2^{(n)}$ with $n>0$. This implies the  subleading
 terms determine the $Q^2$ dependence of $F_{2p}(Q^2)$ at $x=x_0$
and simultaneously the $x$ dependence of the structure functions.
Notice that in the pre-nodal region of 
$Q^2\lesssim 20 \, GeV^2$  the leading and subleading
 structure functions
are very similar in shape.
This explains the failure of the early GLDAP fits: 
only the  limited region of $Q^2 \lesssim 10\, GeV^2$ was
accessible 
and they  could not catch and 
correctly
describe a very different $x$-evolution of 
  pomerons with
different $n$.

At small $x$ only the region $Q^2\lesssim 10^3 \, GeV^2$ is accessible.
In this range the structure functions with $n\geq 3$ 
are hardly distinguishable. Besides, the splitting of the intercept
with $n\geq 3$ is much smaller than for $n=0,1,2$.
 Hence, the Regge expansion (\ref{eq:REGGE})
can be truncated at $n=2$ and  $F_2^{(2)}(Q^2)$ 
comprises contributions from all poles with $n\geq 2$. 

\begin{figure}[b!]
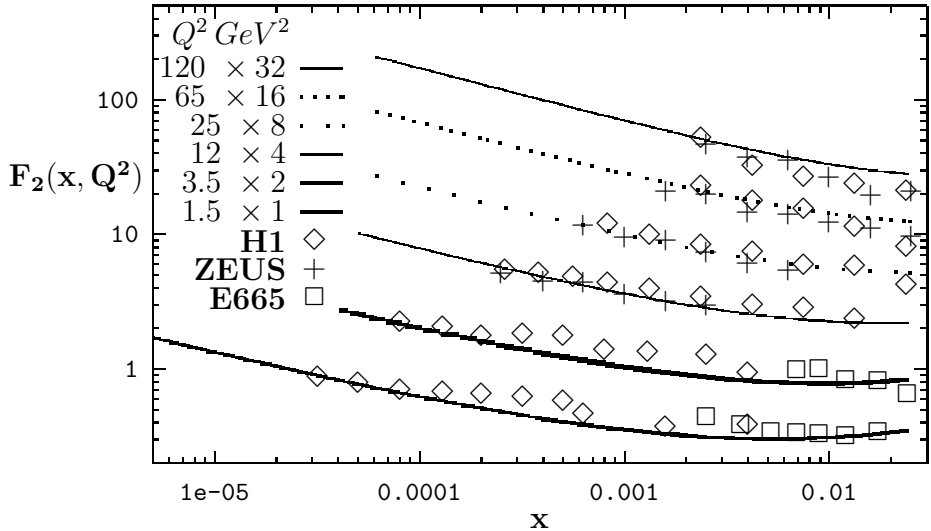
 


\setlength{\unitlength}{0.240900pt}
\ifx\plotpoint\undefined\newsavebox{\plotpoint}\fi
\sbox{\plotpoint}{\rule[-0.200pt]{0.400pt}{0.400pt}}%


\vspace{10pt}
\caption{Three-Pole Approximation vs. H1, ZEUS and E665 data.}
\label{figvrz3}
\end{figure}

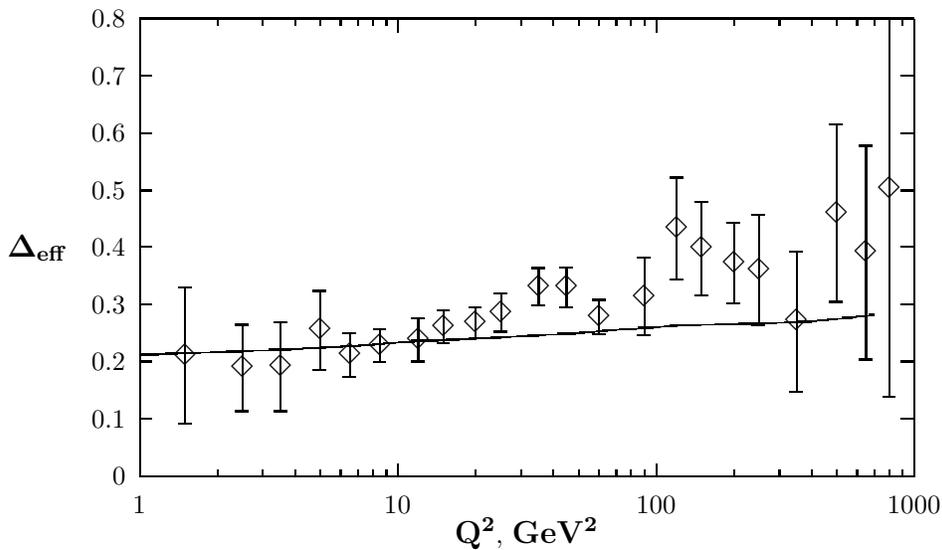
\begin{figure}[b!] 
\setlength{\unitlength}{0.240900pt}
\ifx\plotpoint\undefined\newsavebox{\plotpoint}\fi
\sbox{\plotpoint}{\rule[-0.200pt]{0.400pt}{0.400pt}}%
\begin{picture}(1500,900)(0,0)
\font\gnuplot=cmr10 at 10pt
\gnuplot
\sbox{\plotpoint}{\rule[-0.200pt]{0.400pt}{0.400pt}}%
\put(220.0,113.0){\rule[-0.200pt]{292.934pt}{0.400pt}}
\put(220.0,113.0){\rule[-0.200pt]{4.818pt}{0.400pt}}
\put(198,113){\makebox(0,0)[r]{0}}
\put(1416.0,113.0){\rule[-0.200pt]{4.818pt}{0.400pt}}
\put(220.0,203.0){\rule[-0.200pt]{4.818pt}{0.400pt}}
\put(198,203){\makebox(0,0)[r]{0.1}}
\put(1416.0,203.0){\rule[-0.200pt]{4.818pt}{0.400pt}}
\put(220.0,293.0){\rule[-0.200pt]{4.818pt}{0.400pt}}
\put(198,293){\makebox(0,0)[r]{0.2}}
\put(1416.0,293.0){\rule[-0.200pt]{4.818pt}{0.400pt}}
\put(220.0,383.0){\rule[-0.200pt]{4.818pt}{0.400pt}}
\put(198,383){\makebox(0,0)[r]{0.3}}
\put(1416.0,383.0){\rule[-0.200pt]{4.818pt}{0.400pt}}
\put(220.0,473.0){\rule[-0.200pt]{4.818pt}{0.400pt}}
\put(198,473){\makebox(0,0)[r]{0.4}}
\put(1416.0,473.0){\rule[-0.200pt]{4.818pt}{0.400pt}}
\put(220.0,562.0){\rule[-0.200pt]{4.818pt}{0.400pt}}
\put(198,562){\makebox(0,0)[r]{0.5}}
\put(1416.0,562.0){\rule[-0.200pt]{4.818pt}{0.400pt}}
\put(220.0,652.0){\rule[-0.200pt]{4.818pt}{0.400pt}}
\put(198,652){\makebox(0,0)[r]{0.6}}
\put(1416.0,652.0){\rule[-0.200pt]{4.818pt}{0.400pt}}
\put(220.0,742.0){\rule[-0.200pt]{4.818pt}{0.400pt}}
\put(198,742){\makebox(0,0)[r]{0.7}}
\put(1416.0,742.0){\rule[-0.200pt]{4.818pt}{0.400pt}}
\put(220.0,832.0){\rule[-0.200pt]{4.818pt}{0.400pt}}
\put(198,832){\makebox(0,0)[r]{0.8}}
\put(1416.0,832.0){\rule[-0.200pt]{4.818pt}{0.400pt}}
\put(220.0,113.0){\rule[-0.200pt]{0.400pt}{4.818pt}}
\put(220,68){\makebox(0,0){1}}
\put(220.0,812.0){\rule[-0.200pt]{0.400pt}{4.818pt}}
\put(342.0,113.0){\rule[-0.200pt]{0.400pt}{2.409pt}}
\put(342.0,822.0){\rule[-0.200pt]{0.400pt}{2.409pt}}
\put(413.0,113.0){\rule[-0.200pt]{0.400pt}{2.409pt}}
\put(413.0,822.0){\rule[-0.200pt]{0.400pt}{2.409pt}}
\put(464.0,113.0){\rule[-0.200pt]{0.400pt}{2.409pt}}
\put(464.0,822.0){\rule[-0.200pt]{0.400pt}{2.409pt}}
\put(503.0,113.0){\rule[-0.200pt]{0.400pt}{2.409pt}}
\put(503.0,822.0){\rule[-0.200pt]{0.400pt}{2.409pt}}
\put(535.0,113.0){\rule[-0.200pt]{0.400pt}{2.409pt}}
\put(535.0,822.0){\rule[-0.200pt]{0.400pt}{2.409pt}}
\put(563.0,113.0){\rule[-0.200pt]{0.400pt}{2.409pt}}
\put(563.0,822.0){\rule[-0.200pt]{0.400pt}{2.409pt}}
\put(586.0,113.0){\rule[-0.200pt]{0.400pt}{2.409pt}}
\put(586.0,822.0){\rule[-0.200pt]{0.400pt}{2.409pt}}
\put(607.0,113.0){\rule[-0.200pt]{0.400pt}{2.409pt}}
\put(607.0,822.0){\rule[-0.200pt]{0.400pt}{2.409pt}}
\put(625.0,113.0){\rule[-0.200pt]{0.400pt}{4.818pt}}
\put(625,68){\makebox(0,0){10}}
\put(625.0,812.0){\rule[-0.200pt]{0.400pt}{4.818pt}}
\put(747.0,113.0){\rule[-0.200pt]{0.400pt}{2.409pt}}
\put(747.0,822.0){\rule[-0.200pt]{0.400pt}{2.409pt}}
\put(819.0,113.0){\rule[-0.200pt]{0.400pt}{2.409pt}}
\put(819.0,822.0){\rule[-0.200pt]{0.400pt}{2.409pt}}
\put(869.0,113.0){\rule[-0.200pt]{0.400pt}{2.409pt}}
\put(869.0,822.0){\rule[-0.200pt]{0.400pt}{2.409pt}}
\put(909.0,113.0){\rule[-0.200pt]{0.400pt}{2.409pt}}
\put(909.0,822.0){\rule[-0.200pt]{0.400pt}{2.409pt}}
\put(941.0,113.0){\rule[-0.200pt]{0.400pt}{2.409pt}}
\put(941.0,822.0){\rule[-0.200pt]{0.400pt}{2.409pt}}
\put(968.0,113.0){\rule[-0.200pt]{0.400pt}{2.409pt}}
\put(968.0,822.0){\rule[-0.200pt]{0.400pt}{2.409pt}}
\put(991.0,113.0){\rule[-0.200pt]{0.400pt}{2.409pt}}
\put(991.0,822.0){\rule[-0.200pt]{0.400pt}{2.409pt}}
\put(1012.0,113.0){\rule[-0.200pt]{0.400pt}{2.409pt}}
\put(1012.0,822.0){\rule[-0.200pt]{0.400pt}{2.409pt}}
\put(1031.0,113.0){\rule[-0.200pt]{0.400pt}{4.818pt}}
\put(1031,68){\makebox(0,0){100}}
\put(1031.0,812.0){\rule[-0.200pt]{0.400pt}{4.818pt}}
\put(1153.0,113.0){\rule[-0.200pt]{0.400pt}{2.409pt}}
\put(1153.0,822.0){\rule[-0.200pt]{0.400pt}{2.409pt}}
\put(1224.0,113.0){\rule[-0.200pt]{0.400pt}{2.409pt}}
\put(1224.0,822.0){\rule[-0.200pt]{0.400pt}{2.409pt}}
\put(1275.0,113.0){\rule[-0.200pt]{0.400pt}{2.409pt}}
\put(1275.0,822.0){\rule[-0.200pt]{0.400pt}{2.409pt}}
\put(1314.0,113.0){\rule[-0.200pt]{0.400pt}{2.409pt}}
\put(1314.0,822.0){\rule[-0.200pt]{0.400pt}{2.409pt}}
\put(1346.0,113.0){\rule[-0.200pt]{0.400pt}{2.409pt}}
\put(1346.0,822.0){\rule[-0.200pt]{0.400pt}{2.409pt}}
\put(1373.0,113.0){\rule[-0.200pt]{0.400pt}{2.409pt}}
\put(1373.0,822.0){\rule[-0.200pt]{0.400pt}{2.409pt}}
\put(1397.0,113.0){\rule[-0.200pt]{0.400pt}{2.409pt}}
\put(1397.0,822.0){\rule[-0.200pt]{0.400pt}{2.409pt}}
\put(1417.0,113.0){\rule[-0.200pt]{0.400pt}{2.409pt}}
\put(1417.0,822.0){\rule[-0.200pt]{0.400pt}{2.409pt}}
\put(1436.0,113.0){\rule[-0.200pt]{0.400pt}{4.818pt}}
\put(1436,68){\makebox(0,0){1000}}
\put(1436.0,812.0){\rule[-0.200pt]{0.400pt}{4.818pt}}
\put(220.0,113.0){\rule[-0.200pt]{292.934pt}{0.400pt}}
\put(1436.0,113.0){\rule[-0.200pt]{0.400pt}{173.207pt}}
\put(220.0,832.0){\rule[-0.200pt]{292.934pt}{0.400pt}}
\put(67,472){\makebox(0,0){${\bf \Delta_{eff}}$ }}
\put(828,23){\makebox(0,0){${\bf Q^2}$, ${\bf GeV^2}$}}
\put(828,877){\makebox(0,0){     }}
\put(220.0,113.0){\rule[-0.200pt]{0.400pt}{173.207pt}}
\put(220,304){\usebox{\plotpoint}}
\put(220,304.17){\rule{14.300pt}{0.400pt}}
\multiput(220.00,303.17)(41.320,2.000){2}{\rule{7.150pt}{0.400pt}}
\multiput(291.00,306.61)(19.886,0.447){3}{\rule{12.100pt}{0.108pt}}
\multiput(291.00,305.17)(64.886,3.000){2}{\rule{6.050pt}{0.400pt}}
\multiput(381.00,309.61)(13.188,0.447){3}{\rule{8.100pt}{0.108pt}}
\multiput(381.00,308.17)(43.188,3.000){2}{\rule{4.050pt}{0.400pt}}
\multiput(441.00,312.60)(12.033,0.468){5}{\rule{8.400pt}{0.113pt}}
\multiput(441.00,311.17)(65.565,4.000){2}{\rule{4.200pt}{0.400pt}}
\multiput(524.00,316.61)(12.295,0.447){3}{\rule{7.567pt}{0.108pt}}
\multiput(524.00,315.17)(40.295,3.000){2}{\rule{3.783pt}{0.400pt}}
\multiput(580.00,319.59)(5.832,0.485){11}{\rule{4.500pt}{0.117pt}}
\multiput(580.00,318.17)(67.660,7.000){2}{\rule{2.250pt}{0.400pt}}
\put(657,326.17){\rule{15.500pt}{0.400pt}}
\multiput(657.00,325.17)(44.829,2.000){2}{\rule{7.750pt}{0.400pt}}
\multiput(734.00,328.61)(11.625,0.447){3}{\rule{7.167pt}{0.108pt}}
\multiput(734.00,327.17)(38.125,3.000){2}{\rule{3.583pt}{0.400pt}}
\multiput(787.00,331.61)(12.965,0.447){3}{\rule{7.967pt}{0.108pt}}
\multiput(787.00,330.17)(42.465,3.000){2}{\rule{3.983pt}{0.400pt}}
\multiput(846.00,334.61)(9.616,0.447){3}{\rule{5.967pt}{0.108pt}}
\multiput(846.00,333.17)(31.616,3.000){2}{\rule{2.983pt}{0.400pt}}
\put(890,337.17){\rule{7.100pt}{0.400pt}}
\multiput(890.00,336.17)(20.264,2.000){2}{\rule{3.550pt}{0.400pt}}
\multiput(925.00,339.61)(6.490,0.447){3}{\rule{4.100pt}{0.108pt}}
\multiput(925.00,338.17)(21.490,3.000){2}{\rule{2.050pt}{0.400pt}}
\multiput(955.00,342.60)(8.962,0.468){5}{\rule{6.300pt}{0.113pt}}
\multiput(955.00,341.17)(48.924,4.000){2}{\rule{3.150pt}{0.400pt}}
\multiput(1017.00,346.60)(6.622,0.468){5}{\rule{4.700pt}{0.113pt}}
\multiput(1017.00,345.17)(36.245,4.000){2}{\rule{2.350pt}{0.400pt}}
\multiput(1063.00,350.61)(30.156,0.447){3}{\rule{18.233pt}{0.108pt}}
\multiput(1063.00,349.17)(98.156,3.000){2}{\rule{9.117pt}{0.400pt}}
\multiput(1199.00,353.61)(16.760,0.447){3}{\rule{10.233pt}{0.108pt}}
\multiput(1199.00,352.17)(54.760,3.000){2}{\rule{5.117pt}{0.400pt}}
\multiput(1275.00,356.59)(5.011,0.482){9}{\rule{3.833pt}{0.116pt}}
\multiput(1275.00,355.17)(48.044,6.000){2}{\rule{1.917pt}{0.400pt}}
\multiput(1331.00,362.59)(4.606,0.477){7}{\rule{3.460pt}{0.115pt}}
\multiput(1331.00,361.17)(34.819,5.000){2}{\rule{1.730pt}{0.400pt}}
\put(291,302){\raisebox{-.8pt}{\makebox(0,0){$\Diamond$}}}
\put(381,283){\raisebox{-.8pt}{\makebox(0,0){$\Diamond$}}}
\put(441,285){\raisebox{-.8pt}{\makebox(0,0){$\Diamond$}}}
\put(503,342){\raisebox{-.8pt}{\makebox(0,0){$\Diamond$}}}
\put(550,304){\raisebox{-.8pt}{\makebox(0,0){$\Diamond$}}}
\put(597,318){\raisebox{-.8pt}{\makebox(0,0){$\Diamond$}}}
\put(657,327){\raisebox{-.8pt}{\makebox(0,0){$\Diamond$}}}
\put(697,348){\raisebox{-.8pt}{\makebox(0,0){$\Diamond$}}}
\put(747,354){\raisebox{-.8pt}{\makebox(0,0){$\Diamond$}}}
\put(787,370){\raisebox{-.8pt}{\makebox(0,0){$\Diamond$}}}
\put(846,410){\raisebox{-.8pt}{\makebox(0,0){$\Diamond$}}}
\put(890,410){\raisebox{-.8pt}{\makebox(0,0){$\Diamond$}}}
\put(941,363){\raisebox{-.8pt}{\makebox(0,0){$\Diamond$}}}
\put(1012,395){\raisebox{-.8pt}{\makebox(0,0){$\Diamond$}}}
\put(1063,502){\raisebox{-.8pt}{\makebox(0,0){$\Diamond$}}}
\put(1102,471){\raisebox{-.8pt}{\makebox(0,0){$\Diamond$}}}
\put(1153,447){\raisebox{-.8pt}{\makebox(0,0){$\Diamond$}}}
\put(1192,437){\raisebox{-.8pt}{\makebox(0,0){$\Diamond$}}}
\put(1251,356){\raisebox{-.8pt}{\makebox(0,0){$\Diamond$}}}
\put(1314,526){\raisebox{-.8pt}{\makebox(0,0){$\Diamond$}}}
\put(1360,464){\raisebox{-.8pt}{\makebox(0,0){$\Diamond$}}}
\put(1397,565){\raisebox{-.8pt}{\makebox(0,0){$\Diamond$}}}
\put(291.0,195.0){\rule[-0.200pt]{0.400pt}{51.553pt}}
\put(281.0,195.0){\rule[-0.200pt]{4.818pt}{0.400pt}}
\put(281.0,409.0){\rule[-0.200pt]{4.818pt}{0.400pt}}
\put(381.0,215.0){\rule[-0.200pt]{0.400pt}{32.762pt}}
\put(371.0,215.0){\rule[-0.200pt]{4.818pt}{0.400pt}}
\put(371.0,351.0){\rule[-0.200pt]{4.818pt}{0.400pt}}
\put(441.0,215.0){\rule[-0.200pt]{0.400pt}{33.726pt}}
\put(431.0,215.0){\rule[-0.200pt]{4.818pt}{0.400pt}}
\put(431.0,355.0){\rule[-0.200pt]{4.818pt}{0.400pt}}
\put(503.0,280.0){\rule[-0.200pt]{0.400pt}{29.872pt}}
\put(493.0,280.0){\rule[-0.200pt]{4.818pt}{0.400pt}}
\put(493.0,404.0){\rule[-0.200pt]{4.818pt}{0.400pt}}
\put(550.0,269.0){\rule[-0.200pt]{0.400pt}{16.622pt}}
\put(540.0,269.0){\rule[-0.200pt]{4.818pt}{0.400pt}}
\put(540.0,338.0){\rule[-0.200pt]{4.818pt}{0.400pt}}
\put(597.0,292.0){\rule[-0.200pt]{0.400pt}{12.527pt}}
\put(587.0,292.0){\rule[-0.200pt]{4.818pt}{0.400pt}}
\put(587.0,344.0){\rule[-0.200pt]{4.818pt}{0.400pt}}
\put(657.0,293.0){\rule[-0.200pt]{0.400pt}{16.381pt}}
\put(647.0,293.0){\rule[-0.200pt]{4.818pt}{0.400pt}}
\put(647.0,361.0){\rule[-0.200pt]{4.818pt}{0.400pt}}
\put(697.0,322.0){\rule[-0.200pt]{0.400pt}{12.286pt}}
\put(687.0,322.0){\rule[-0.200pt]{4.818pt}{0.400pt}}
\put(687.0,373.0){\rule[-0.200pt]{4.818pt}{0.400pt}}
\put(747.0,330.0){\rule[-0.200pt]{0.400pt}{11.563pt}}
\put(737.0,330.0){\rule[-0.200pt]{4.818pt}{0.400pt}}
\put(737.0,378.0){\rule[-0.200pt]{4.818pt}{0.400pt}}
\put(787.0,340.0){\rule[-0.200pt]{0.400pt}{14.454pt}}
\put(777.0,340.0){\rule[-0.200pt]{4.818pt}{0.400pt}}
\put(777.0,400.0){\rule[-0.200pt]{4.818pt}{0.400pt}}
\put(846.0,381.0){\rule[-0.200pt]{0.400pt}{14.213pt}}
\put(836.0,381.0){\rule[-0.200pt]{4.818pt}{0.400pt}}
\put(836.0,440.0){\rule[-0.200pt]{4.818pt}{0.400pt}}
\put(890.0,378.0){\rule[-0.200pt]{0.400pt}{15.177pt}}
\put(880.0,378.0){\rule[-0.200pt]{4.818pt}{0.400pt}}
\put(880.0,441.0){\rule[-0.200pt]{4.818pt}{0.400pt}}
\put(941.0,336.0){\rule[-0.200pt]{0.400pt}{13.009pt}}
\put(931.0,336.0){\rule[-0.200pt]{4.818pt}{0.400pt}}
\put(931.0,390.0){\rule[-0.200pt]{4.818pt}{0.400pt}}
\put(1012.0,334.0){\rule[-0.200pt]{0.400pt}{29.390pt}}
\put(1002.0,334.0){\rule[-0.200pt]{4.818pt}{0.400pt}}
\put(1002.0,456.0){\rule[-0.200pt]{4.818pt}{0.400pt}}
\put(1063.0,422.0){\rule[-0.200pt]{0.400pt}{38.544pt}}
\put(1053.0,422.0){\rule[-0.200pt]{4.818pt}{0.400pt}}
\put(1053.0,582.0){\rule[-0.200pt]{4.818pt}{0.400pt}}
\put(1102.0,397.0){\rule[-0.200pt]{0.400pt}{35.412pt}}
\put(1092.0,397.0){\rule[-0.200pt]{4.818pt}{0.400pt}}
\put(1092.0,544.0){\rule[-0.200pt]{4.818pt}{0.400pt}}
\put(1153.0,384.0){\rule[-0.200pt]{0.400pt}{30.594pt}}
\put(1143.0,384.0){\rule[-0.200pt]{4.818pt}{0.400pt}}
\put(1143.0,511.0){\rule[-0.200pt]{4.818pt}{0.400pt}}
\put(1192.0,350.0){\rule[-0.200pt]{0.400pt}{41.676pt}}
\put(1182.0,350.0){\rule[-0.200pt]{4.818pt}{0.400pt}}
\put(1182.0,523.0){\rule[-0.200pt]{4.818pt}{0.400pt}}
\put(1251.0,245.0){\rule[-0.200pt]{0.400pt}{53.239pt}}
\put(1241.0,245.0){\rule[-0.200pt]{4.818pt}{0.400pt}}
\put(1241.0,466.0){\rule[-0.200pt]{4.818pt}{0.400pt}}
\put(1314.0,387.0){\rule[-0.200pt]{0.400pt}{67.211pt}}
\put(1304.0,387.0){\rule[-0.200pt]{4.818pt}{0.400pt}}
\put(1304.0,666.0){\rule[-0.200pt]{4.818pt}{0.400pt}}
\put(1360.0,296.0){\rule[-0.200pt]{0.400pt}{80.942pt}}
\put(1350.0,296.0){\rule[-0.200pt]{4.818pt}{0.400pt}}
\put(1350.0,632.0){\rule[-0.200pt]{4.818pt}{0.400pt}}
\put(1397.0,238.0){\rule[-0.200pt]{0.400pt}{143.095pt}}
\put(1387.0,238.0){\rule[-0.200pt]{4.818pt}{0.400pt}}
\put(1387.0,832.0){\rule[-0.200pt]{4.818pt}{0.400pt}}
\end{picture}

\vspace{10pt}
\caption{Effective intercept vs. H1 data.}
\label{figvrz4}
\end{figure}

The BFKL equation  allows one
to determine  the intercepts and structure functions $F_2^{(n)}(Q^2)$.
The only adjustable parameters are "the pole residues"
$A_0,\,,A_1\,,A_2$ in (\ref{eq:REGGE}) which are fixed, in fact,
by the boundary condition at $x=x_0$.
With the  proper account of the  valence \cite{GLUCK}
and non-perturbative \cite{PLNZ2} corrections to (\ref{eq:REGGE}) we arrive at 
the three-pole-approximation
 which appears to be very successful when confronted with
the data \cite{DATH1}. 
in a wide kinematical range 
(Figure \ref{figvrz3}). 
The effective pomeron intercept 
\begin{equation}
\Delta_{eff}=-{\partial \log F_{2p}(x,Q^2)\over \partial \log x}
 \label{eq:DELEFF}
\end{equation}
gives an idea of  the role of the subleading singularities.
The intercept $\Delta_{eff}$ 
  calculated with  the experimental kinematic constraints 
 is much smaller than $\Delta_0=0.4$
which is expected to dominate asymptotically. 
 The agreement of our numerical estimates
 with the $H1$ determination (Figure \ref{figvrz4})
is quite satisfactory.

{\bf Acknowledgments} I would like to thank Jos\'e Repond 
for the kind hospitality in Chicago and careful reading the 
manuscript. 
 
\end{document}